\documentclass{svproc}

\usepackage[english]{babel}
\usepackage{comment}

\usepackage{amsmath}
\usepackage{graphicx}
\usepackage{booktabs}
\usepackage{tabularx}
\usepackage{xurl}

\usepackage{xcolor}
\definecolor{darkred}{rgb}{0.6,0.0,0.0}
\definecolor{darkgreen}{rgb}{0,0.50,0}
\definecolor{lightblue}{rgb}{0.0,0.42,0.91}
\definecolor{orange}{rgb}{0.99,0.48,0.13}
\definecolor{grass}{rgb}{0.18,0.80,0.18}
\definecolor{pink}{rgb}{0.97,0.15,0.45}

\usepackage{listings}

\usepackage[numbers]{natbib}
\usepackage[colorlinks=true,allcolors=blue]{hyperref}

\newcommand{\orcidlink}[1]{\href{https://orcid.org/#1}{\includegraphics[scale=0.06]{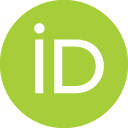}}}

\title{A Methodology for Analysing Coding Bugs for LLMs and SAST}

\author{
    Madjid G. Tehrani\,\orcidlink{0000-0002-4838-5865}\inst{1},
    William J. Buchanan\,\orcidlink{0000-0003-0809-3523}\inst{1},
    Eldar Sultanow\,\orcidlink{0000-0001-5257-2236}\inst{3},
    Mahkame Houmani\,\orcidlink{0009-0005-2708-37}\inst{2},
    Christel H. Djaha Fodja\,\orcidlink{0009-0009-3211-0128}\inst{2},
    Mouad Lemoudden\inst{1}
}

\institute{
Blockpass ID Lab, Edinburgh Napier University, Edinburgh, UK
\and
The George Washington University, Washington DC, USA
\and
Capgemini Deutschland GmbH, Nuremberg, Germany
}

\authorrunning{Madjid G. Tehrani et al.}

\begin{document}
\maketitle

\begin{abstract}
 Large language models (LLMs) are increasingly used for code understanding, yet their practical effectiveness for vulnerability detection relative to Static Application Security Testing (SAST) remains insufficiently quantified. We present a controlled comparative study between GPT-4 (Advanced Data Analysis) and two SAST tools (SonarQube and Cloud Defence) on 32 curated security scenarios representing common coding pitfalls. Each scenario is scored with a binary detection rule, the two SAST outputs are aggregated using a logical OR baseline, and paired outcomes are evaluated using McNemar's test for statistical significance. In our dataset, GPT-4 correctly detected 30 of 32 scenarios (93.75\%), while the aggregated SAST baseline detected 11 of 32. The paired comparison shows a statistically significant difference in detection performance in favour of GPT-4. We also discuss security considerations and operational constraints for integrating LLM-enhanced vulnerability scanning into secure software development workflows.
 
\keywords{Large language models, static application security testing, vulnerability detection, McNemar's test}

\end{abstract}

\section{Introduction}
One of the most significant threats within cybersecurity is the application of zero-day threats. These tend to be caused by unforeseen vulnerabilities in code and which can open up significant weaknesses within our society. For example, Heartbleed \cite{ghafoor2014analysis} was one significant example of this, and where the OpenSSL software package allowed an adversary to capture part of the running memory on a server, and which could reveal passwords and cryptographic keys. And, so, while many developers are diligent in their approach, there are still many cases of code being released without a strong focus on cybersecurity testing. A well-defined trait is that software developers often have an eye for usability and in producing workable code rather than for it to be safe from future vulnerabilities \cite{sen2018challenges}. 

This paper aims to understand if machine-learning-aided testing for cybersecurity could enhance our existing static approaches to software testing. A core contribution of the paper is in the testing of the accuracy performance of the machine-learning approach of GPT-4 against the usage of static testing tools, along with a well-defined methodology that can be used in the assessment for the accuracy of bug finding. It uses a sampling-based approach to code examples and focuses on well-defined classifications of software bugs as defined within the MITRE Common Weakness Enumeration (CWE).

The remainder of this paper is organised as follows. Section 2 reviews related work on LLM-assisted bug finding and vulnerability detection. Section 3 presents the methodology, including the experiment design and the security scenarios used for evaluation. Section 4 reports the comparative results between GPT-4 (Advanced Data Analysis) and the selected SAST tools. Section 5 discusses the implications of the findings, outlines key security concerns for LLM-enhanced SAST, and identifies directions for future research. Section 6 concludes the paper and summarises the main limitations of the study. Section 7 provides an appendix with the statistical testing script used in the evaluation.

\subsection{Background}

Machine learning for code analysis in cybersecurity has been widely studied \cite{sharma2021survey}, and recent advances in natural language processing (NLP) have enabled large language models (LLMs) such as GPT-4 \cite{openai_2023_gpt4}. In practice, however, vulnerability discovery still relies heavily on Static Application Security Testing (SAST), which uses static code analysis to flag potential security defects prior to deployment. A persistent problem is that SAST tools often miss exploitable vulnerabilities (false negatives) in diverse coding patterns, while also producing findings that developers may deprioritise due to false positives. This slows remediation and widens the window of vulnerability, defined as the time between exploit availability and patch deployment \cite{johansen_2007_firepatch}. In parallel, operational and policy constraints can delay disclosure and verification of zero-days, further increasing systemic exposure \cite{CSL,_2022_tianfu,_2023_dsip,_2023_ai,_2023_machine}.

The research gap addressed in this paper is the lack of a transparent, reproducible methodology for measuring whether a general-purpose LLM can detect a broad set of real-world vulnerability patterns more effectively than widely used SAST tools, using clearly defined scenarios and statistically grounded comparison. Our focus is not to replace SAST, but to quantify detection capability under controlled conditions and to characterise when an LLM is complementary to static analysis.

We consider two common classes of application security testing:
\begin{itemize}
    \item Static Application Security Testing (SAST): automated analysis of source code in a non-runtime setting to detect security-relevant defects and insecure coding patterns.
    \item Dynamic Application Security Testing (DAST): analysis of an application during execution, exercising program paths and inputs in a runtime environment.
\end{itemize}

\section{Related Work}
Dominik Sobania et al. \cite{sobania_2023_an} explored automated program repair techniques, specifically focusing on ChatGPT's potential for bug fixing. According to them, while initially not designed for this purpose, ChatGPT demonstrated promising results on the QuixBugs benchmark, rivalling advanced methods like CoCoNut and Codex. ChatGPT's interactive dialogue system uniquely enhances its repair rate, outperforming established standards.
Wei Ma et al. \cite{ma_2023} noted that while ChatGPT shows impressive potential in software engineering(SE) tasks like code and document generation, its lack of interpretability raises concerns given SE's high-reliability requirements. Through a detailed study, they categorised AI's essential skills for SE into syntax understanding, static behaviour understanding, and dynamic behaviour understanding. Their assessment, spanning languages like C, Java, Python, and Solidity, revealed that ChatGPT excels in syntax understanding (akin to an AST parser) but faces challenges in comprehending dynamic semantics. The study also found ChatGPT prone to hallucinations, emphasising the need to validate its outputs for SE dependability and suggesting that codes from LLMs are syntactically correct but potentially vulnerable.

Haonan Li et al. \cite{li_2023} discussed the challenges of balancing precision and scalability in static analysis for identifying software bugs. While LLMs show potential in understanding and debugging code, their efficacy in handling complex bug logic, which often requires intricate reasoning and broad analysis, remains limited. Therefore, the researchers suggest using LLMs to assist rather than replace static analysis. Their study introduced LLift, an automated system combining a static analysis tool and an LLM to address use-before-initialisation (UBI) bugs. Despite various challenges like bug-specific modelling and the unpredictability of LLMs, LLift, when tested on real-world potential UBI bugs, showed significant precision (50\%) and recall (100\%). Notably, it uncovered 13 new UBI bugs in the Linux kernel, highlighting the potential of LLM-assisted methods in extensive real-world bug detection.

Norbert Tihani et al. \cite{tihanyi_2023} introduced the FormAI dataset, comprising 112,000 AI-generated C programs with vulnerability classifications generated by GPT-3.5-turbo. These programs range from complex tasks like network management and encryption to simpler ones, like string operations. Each program comes labelled with the identified vulnerabilities, pinpointing the type, line number, and vulnerable function. To achieve accurate vulnerability detection without false positives, the Efficient SMT-based Bounded Model Checker (ESBMC) was used. This method leverages techniques like model checking and constraint programming to reason over program safety. Each vulnerability also references its corresponding Common Weakness Enumeration (CWE) number. 

Codex, introduced by Mark et al. \cite{chen_2021_evaluating}, represents a significant advancement in GPT language models, tailored specifically for code synthesis using data from GitHub. This refined model underpins the operations of GitHub Copilot. When assessed on the HumanEval dataset, designed to gauge the functional accuracy of generating programs based on docstrings, Codex achieved a remarkable 28.8\% success rate. In stark contrast, GPT-3 yielded a 0\% success rate, and GPT-J achieved 11.4\%. A standout discovery was the model's enhanced performance through repeated sampling, with a success rate soaring to 70.2\% when given 100 samples per problem. Despite these promising results, Codex does exhibit certain limitations, notably struggling with intricate docstrings and variable binding operations. The paper deliberates on the broader ramifications of deploying such potent code-generation tools, touching upon safety, security, and economic implications.

In a technical evaluation, Cheshkov et al. \cite{cheshkov_2023_technical} found that the ChatGPT and GPT-3 models, despite their success in various other code-based tasks, performed on par with a dummy classifier for this particular challenge. Utilising a dataset of Java files sourced from GitHub repositories, the study emphasised the models' current limitations in the domain of vulnerability detection. However, the authors remain optimistic about the potential of future advancements, suggesting that models like GPT-4, with targeted research, could eventually make significant contributions to the field of vulnerability detection.

A comprehensive study conducted by Xin Liu et al. \cite{liu_2023_not} investigated the potential of ChatGPT in Vulnerability Description Mapping (VDM) tasks. VDM is pivotal in efficiently mapping vulnerabilities to CWE and MITRE ATT\&CK Techniques classifications. Their findings suggest that while ChatGPT approaches the proficiency of human experts in the Vulnerability-to-CWE task, especially with high-quality public data, its performance is notably compromised in tasks such as Vulnerability-to-ATT\&CK, particularly when reliant on suboptimal public data quality. Ultimately, Xin Liu et al. emphasise that, despite the promise shown by ChatGPT, it is not yet poised to replace the critical expertise of professional security engineers, asserting that closed-source LLMs are not the conclusive answer for VDM tasks.
Last but not least, the OWASP top 10 for LLMs \cite{a2023_owasp} introduced ten security risks as follows: Prompt Injection, Insecure Output Handling, Training Data Poisoning, Model Denial of Service, Supply Chain Vulnerabilities, Sensitive Information Disclosure, Insecure Plugin Design, Excessive Agency, Over reliance, and Model Theft.

In contrast to prior work that focuses on program repair, synthetic datasets, or vulnerability mapping tasks, this study targets an end-to-end and reproducible comparison of an LLM and SAST on a curated set of security pitfalls with paired statistical testing. Our contribution is not a new detector, but a transparent evaluation protocol (scenario selection, binary scoring, aggregated SAST baseline, and McNemar inference) that enables controlled measurement of relative detection capability under identical inputs.

\section{Methodology and security scenarios}
\subsection{Experiment Design and Data}

We selected two different Static Application Security Testing (SAST) tools to ensure a fair comparison. The first is SonarCloud\footnote{\url{https://www.sonarsource.com/products/sonarcloud/}}, a well-established SAST platform provided by SonarSource. SonarQube supports 29 languages and offers continuous inspection of code quality. It conducts automatic reviews through static analysis to detect bugs, code smells, and other issues across the 29 supported languages. The platform provides insights on duplicated code, coding standards, unit tests, code coverage, code complexity, comments, bugs, and security vulnerabilities.

On the other hand, the second tool is a relatively new, paid Software-as-a-Service (SaaS) that began operations in 2020, named Cloud Defence\footnote{\url{https://www.clouddefense.ai/}}. Its mission is "to shield Cloud Native Applications from Zero Day Attacks." To derive a comprehensive evaluation, we combined the outcomes of both tools using an 'OR' operation and named this consolidated result the "SAST result." Consequently, a positive outcome (indicated by a '1') in the SAST result signifies that either of the tools successfully detected the security vulnerability.

For this study, we selected 32 known security pitfalls that developers might inadvertently introduce, potentially leading to zero-day vulnerabilities. Based on our observations, we formulated the following hypotheses:
\begin{itemize}
    \item \textbf{\(H_0\)}: GPT-4-Advanced Data Analysis detects vulnerabilities with the same or worse performance than the SAST tools.
    \item \textbf{\(H_1\)}: GPT-4-Advanced Data Analysis detects vulnerabilities with better performance than the SAST tools.
\end{itemize}

\textbf{Design of the Experiment}
We have taken the code samples from GitHub or Snyk. Each sample corresponds to a known security pitfall listed in Table~\ref{table:scenarios} and is evaluated against its intended vulnerability class (reported with a CWE identifier in Table~\ref{table:results}). We ran each code sample independently through the GPT-4-Advanced Data Analysis web interface and through the selected SAST tools.The web interface does not expose the exact context window and system configuration used during analysis, which we treat as a study limitation; the prompts and outputs used for scoring are included in the replication package (see footnote).
Detailed records of prompts, tool configurations as applicable, tool outputs, and scoring decisions were maintained and are accessible \footnote{\url{https://github.com/owlmt/owmlt-llm-sast-artifacts}}.

To enhance clarity and transparency, we applied the same end-to-end procedure to every sample:
\begin{enumerate}
    \item \textbf{Sample selection and labelling:} select a representative snippet that exhibits the intended vulnerability pattern; label the sample with its intended vulnerability category and corresponding CWE reference used for reporting.
    \item \textbf{SAST execution:} analyse the snippet with SonarQube and Cloud Defence using their standard workflows; record whether each tool flags the intended vulnerability category or an equivalent security finding consistent with the assigned label.
    \item \textbf{LLM execution:} submit the same snippet to GPT-4-Advanced Data Analysis with a fixed analysis instruction to identify security vulnerabilities and explain the issue; record the full response.
    \item \textbf{Binary scoring:} score each detector output as 1 for correct detection and 0 for a miss, using the scoring rule defined below.
    \item \textbf{SAST aggregation:} compute an aggregated SAST result using a logical OR across the two SAST tools; the aggregated SAST outcome is 1 if either tool detects the vulnerability, otherwise 0.
\end{enumerate}

\textbf{Scoring rule}
For each sample, a detector was scored as correct (1) if it explicitly identified the intended vulnerability type for that sample in a way consistent with the assigned category (for example, SQL injection, XSS, SSRF, insecure deserialization, directory traversal). If the detector produced no relevant finding, flagged only unrelated issues, or described a different vulnerability class, the outcome was scored as incorrect (0). Where a detector reported a closely related label, the decision was resolved by checking whether the described weakness matches the vulnerability mechanism in the code and aligns with the assigned CWE used in Table~\ref{table:results}. Evidence supporting each scoring decision is included in the replication package (see footnote above).

For each tool's detection ability, outcomes were categorised binarily: 1 denoting correct detection and 0 indicating a miss. Given the comparative nature of the study, the Chi-Squared Test for Independence was chosen as an initial framing to compare detection outcomes. However, because each code sample yields paired binary outcomes (GPT-4 and aggregated SAST on the same sample), we used McNemar's test \cite{mcnemar_1947_note} for statistical inference on paired nominal data. We made a 2x2 contingency table that was formulated as follows:

\begin{center}
\begin{tabular}{|c|c|c|}
\hline
                   & GPT-4 Correct & GPT-4 Incorrect \\
\hline
SAST Correct      & a             & b               \\
\hline
SAST Incorrect    & c             & d               \\
\hline
\end{tabular}
\end{center}

Wherein:
\begin{itemize}
    \item \(a\): Represents vulnerabilities correctly identified by both tools.
    \item \(b\): Represents vulnerabilities exclusively detected by the SAST tool.
    \item \(c\): Represents vulnerabilities exclusively identified by GPT-4.
    \item \(d\): Represents vulnerabilities that remained undetected by both entities.
\end{itemize}

\textbf{Interpretation of Outcomes}
We used McNemar's test \cite{mcnemar_1947_note}, and the p-value below 0.05 was established as the benchmark for statistical significance. If attained, it would signify a superior performance of GPT-4 over the SAST tool in the scope of vulnerability detection.

\subsection{Security Vulnerabilities and Data}
We explored the security scenarios that are listed in Table~\ref{table:scenarios} and examined GPT4 and a SAST for each. Code snippets, detector outputs, and supporting evidence for the scoring decisions are documented in the replication package (see footnote above).

\begin{table}

\centering
\caption{Brief Descriptions of Various Attacks and sample code}
\begin{tabular}{p{0.4cm} p{4.4 cm} p{1.2cm} p{9cm}}
\toprule
\textbf{ID} & \textbf{Coding security mistakes} & \textbf{sample}& \textbf{Description} \\
\midrule
1 & Buffer overflow & \cite{a2023_buffer_overflowc} & Overwriting memory by overflowing a buffer. \\
\hline
2 & SQL Injection & \cite{a2019_sqlinjectionattack, SQLi} & injecting malicious SQL code into a query. \\
\hline
3 & Cross-Site Scripting (XSS) & \cite{xss-c} & Injecting malicious scripts into web pages viewed by users. \\
\hline
4 & Broken Access Control & \cite{a2022_broken} & Improperly enforcing what users can or cannot do. \\
\hline
5 & Insecure deserialization & \cite{Deserialization} & Exploiting unsafe data unmarshalling. \\
\hline
6 & Log4J & \cite{log4j} & Exploiting the Log4J Java logging library. \\
\hline
7 & Unrestricted upload & \cite{unrestricted} & Uploading malicious files without restrictions. \\
\hline
8 & Improper input validation & \cite{validation} & Not verifying the user's input properly. \\
\hline
9 & Memory Leak & \cite{memory} & Unintentional memory consumption leading to crashes. \\
\hline
10 & Mass assignment & \cite{mass} & Overwriting object properties without restrictions. \\
\hline
11 & Server-side request forgery & \cite{SSRF} & Making the server run unauthorized actions. \\
\hline
12 & Insecure temporary files & \cite{temporary} & Exploiting insecurely created temporary files. \\
\hline
13 & Cleartext storage in a cookie & \cite{cookie} & Storing sensitive data unencrypted in cookies. \\
\hline
14 & XPath injection & \cite{XPath} & Injecting malicious XPath queries. \\
\hline
15 & Weak password recovery & \cite{password} & Exploiting inadequate password recovery systems. \\
\hline
16 & Logging vulnerabilities & \cite{logging} & Inadequately protecting or revealing logs. \\
\hline
17 & Insecure Randomness & \cite{randomness} & Using predictable random number generators. \\
\hline
18 & NoSQL injection attack & \cite{nosql} & Injecting malicious code into NoSQL queries. \\ \hline
19 & Code injection & \cite{code} & Injecting malicious code into an application. \\
\hline
20 & No rate limiting & \cite{Rate} & Overloading systems by not capping request rates. \\
\hline
21 & Vulnerable components & \cite{Outdated} & Using outdated or flawed software components. \\
\hline
22 & Insecure design & \cite{Design} & Designing systems without security in mind. \\
\hline
23 & Insecure hash & \cite{hash} & Using weak hashing methods. \\
\hline
24 & ReDoS & \cite{ReDoS} & Exploiting regex to cause denial-of-service. \\ \hline
25 & XML external entity injection & \cite{XXE} & Attacking parsers with external XML entities. \\
\hline
26 & Cross site request forgery & \cite{csrf} & Making users unknowingly submit a malicious request. \\
\hline
27 & DOM XSS & \cite{dom} & Injecting malicious scripts via the Document Object Model. \\
\hline
28 & Open redirect & \cite{redirect} & Redirecting users to malicious sites. \\
\hline
29 & Directory traversal & \cite{directory} & Accessing files outside of the intended directory. \\
\hline
30 & Prototype pollution & \cite{prototype} & Altering prototype objects. \\
\hline
31 & Container capabilities & \cite{capabilities_container} & Containers retaining unnecessary capabilities. \\
\hline
32 & Container privileged mode & \cite{privileged_container} & Running containers with full system privileges. \\ \hline
\bottomrule

\end{tabular}

\label{table:scenarios}

\end{table}

\section{Results}
The comparison results between the online SAST tool and GPT-4 for detecting security vulnerabilities are presented in Table~\ref{table:results}. Vulnerabilities ranged from common issues like Buffer Overflow and SQL Injection to more specific ones like Prototype Pollution. GPT-4 consistently detected most vulnerabilities correctly, as indicated by a "1" under the "GPT-C" column. In contrast, tools like sonarcloud and clouddefense had varied results, with some vulnerabilities detected correctly and others not.
Across 32 paired scenarios, the aggregated SAST baseline detected 11 cases (34.38\%), while GPT-4 detected 30 cases (93.75\%), an absolute difference of 59.37 percentage points. The paired outcomes in Table~\ref{table:comparison} show \(b=0\) (SAST-only detections) and \(c=22\) (GPT-4-only detections), indicating that most disagreements arise from SAST misses where GPT-4 flags the intended vulnerability. This directional imbalance explains the strong McNemar significance reported below.

\begin{table}
\small
\centering

\caption{Comparison results for various security vulnerabilities. C: Correct detection; I: Incorrect detection. Tools compared include GPT-4 Advanced Data Analysis (GPT), sonarcloud.io (SQ), and clouddefense.ai (CDA). Vulnerabilities are referenced by their Common Weakness Enumeration (CWE) ID, available at \url{https://cwe.mitre.org/}.}
\begin{tabular}{p{1cm} p{6.9cm} p{1.0cm} p{1.0cm} p{1.0cm} p{1.0cm} p{1.0cm} p{1.0cm} p{1.0cm} p{1cm}}

\toprule 
ID & Security vulnerability inside a code snippet with its CWE & GPT-C & GPT-I & SQ-C & SQ-I & CDA-C & CDA-I & CWE \\
\midrule
1 & Buffer overflow \cite{cwe121} & 1 & 0 & 1 & 0 & 1 & 0 & 121 \\ \hline
2 & SQL Injection \cite{cwe564}& 1 & 0 & 1 & 0 & 1 & 0 & 564 \\ \hline
3 & Cross-Site Scripting (XSS) \cite{cwe79}: & 1 & 0 & 0 & 1 & 0 & 1 & 79 \\ \hline
4 & Broken Access Control \cite{cwe284} & 1 & 0 & 0 & 1 & 0 & 1 & 284 \\ \hline
5 & Insecure deserialization \cite{cwe502} & 1 & 0 & 0 & 1 & 0 & 1 & 502 \\ \hline
6 & $\log 4 J$ \cite{cwe502}& 1 & 0 & 0 & 1 & 0 & 1 & 502 \\ \hline
7 & Unrestricted upload of dangerous files \cite{cwe434}& 1 & 0 & 0 & 1 & 0 & 1 & 434 \\ \hline
8 & Improper input validation \cite{cwe20}& 1 & 0 & 0 & 1 & 0 & 1 & 20 \\ \hline
9 & Memory Leak \cite{cwe401} & 1 & 0 & 0 & 1 & 1 & 0 & 401 \\ \hline
10 & Mass assignment with secret leak \cite{cwe915} & 1 & 0 & 0 & 1 & 0 & 1 & 915 \\ \hline
11 & Server-side request forgery \cite{cwe918} & 1 & 0 & 1 & 0 & 1 & 0 & 918 \\ \hline
12 & Insecure temporary file \cite{cwe377}& 1 & 0 & 0 & 1 & 0 & 1 & 377 \\ \hline
13 & Plaintext storage of sensitive information in cookies \cite{cwe315} & 1 & 0 & 0 & 1 & 0 & 1 & 315 \\ \hline
14 & XPath injection \cite{cwe643} & 1 & 0 & 0 & 1 & 0 & 1 & 643 \\ \hline
15 & Weak password recovery \cite{cwe640} & 1 & 0 & 0 & 1 & 0 & 1 & 640 \\ \hline
16 & Logging vulnerabilities \cite{cwe532} & 1 & 0 & 0 & 1 & 0 & 1 & 532 \\ \hline 
17 & Insecure Randomness \cite{cwe330} & 1 & 0 & 0 & 1 & 0 & 1 & 330 \\ \hline
18 & NoSQL injection attack \cite{cwe89}& 1 & 0 & 0 & 1 & 0 & 1 & 89 \\ \hline
19 & Code injection \cite{cwe94}& 1 & 0 & 1 & 0 & 1 & 0 & 94 \\ \hline
20 & No rate limiting \cite{cwe770}& 1 & 0 & 0 & 1 & 0 & 1 & 770 \\ \hline
21 & Vulnerable and outdated components \cite{cwe1352}& 0 & 1 & 0 & 1 & 0 & 1 & 1352 \\ \hline
22 & Insecure design \cite{cwe657} & 0 & 1 & 0 & 1 & 0 & 1 & 657 \\ \hline
23 & Insecure hash \cite{cwe328}& 1 & 0 & 0 & 1 & 0 & 1 & 328 \\ \hline
24 & $\operatorname{ReDoS}$ \cite{cwe185}& 1 & 0 & 0 & 1 & 0 & 1 & 185 \\ \hline
25 & XML external entity injection \cite{cwe611}& 1 & 0 & 1 & 0 & 1 & 0 & 611 \\ \hline
26 & Cross-site request forgery \cite{cwe352} & 1 & 0 & 0 & 1 & 0 & 1 & 352 \\ \hline
27 & DOM XSS \cite{cwe80} & 1 & 0 & 1 & 0 & 1 & 0 & 80 \\ \hline
28 & Open redirect \cite{cwe601}& 1 & 0 & 1 & 0 & 1 & 0 & 601 \\ \hline
29 & Directory traversal \cite{cwe23} & 1 & 0 & 0 & 1 & 1 & 0 & 23 \\ \hline
30 & Prototype pollution \cite{cwe1321} & 1 & 0 & 0 & 1 & 0 & 1 & 1321 \\ \hline
31 & Container does not drop default capabilities \cite{cwe250} & 1 & 0 & 0 & 1 & 0 & 1 & 250 \\ \hline
32 & Container is running in privileged mode \cite{cwe250}& 1 & 0 & 0 & 1 & 0 & 1 & 250 \\ \hline
\bottomrule
\end{tabular}

\label{table:results}
\end{table}

To enhance the generalizability of our method to encompass a broader range of SAST tools, we introduced new columns named SAST-Correct and SAST-Incorrect into Table~\ref{table:comparison}. We constructed the contingency matrix using these columns and GPT-Correct and GPT-Incorrect.

$$
S A S T_{\text {correct }}=\bigcup_{S A S T} \operatorname{Result}_{S A S T}^{\text {correct }} ; S A S T_{\text {incorrect }}=1-S A S T_{\text {correct }}
$$

\begin{table}[!ht]
\centering

\begin{tabular}{|c|c|c|}
\hline
 & GPT-4 Correct & GPT-4 Incorrect \\
\hline
SAST Correct & 11 & 0 \\
\hline
SAST Incorrect & 22 & 2 \\
\hline
\end{tabular}
\caption{Comparison of SAST and GPT-4 Detection Abilities}
\label{table:comparison}
\end{table}
Utilising McNemar's test 
\cite{mcnemar_1947_note}, a comparative evaluation of vulnerability detection performance between GPT-4 and SAST tools was conducted. The test yielded a Chi-square value of 20.046 with an associated p-value of 0.000007562  Algorithm in the Appendix.

Given this result and adopting a significance level of 5\% (i.e., $\alpha = 0.05$), we reject the null hypothesis, and our experiment supports the alternative hypothesis:

\begin{itemize}
    \item \(H_0\): GPT-4-Advanced Data Analysis has the same or worse performance than SAST tools.
    \item \(H_1\): GPT-4-Advanced Data Analysis has better performance than SAST tools.
\end{itemize}

\section{Discussion and future research}
GPT-4 has shown a promising ability to detect vulnerabilities that traditional SAST tools might miss. This revelation is significant for several reasons:

\begin{itemize}
    \item Evolution of Detection Tools: As software development processes evolve, so too must the tools that ensure their security. The capabilities of GPT-4 in our experiment suggest that language models can serve as powerful supplements, if not alternatives, to traditional SAST tools.
    \item Cost Implications: Traditional SAST tools, especially proprietary ones, can be expensive. If language models can provide comparable or even superior performance, organisations might be able to reduce costs associated with security testing.
    \item Time Efficiency: The rapid analysis capabilities of models like GPT-4 could reduce the time taken for security assessments, especially in continuous integration/continuous deployment (CI/CD) environments.
\end{itemize}

However, while GPT-4's performance is commendable, it is essential to approach these findings with caution.
Language models, no matter how advanced, are not infallible. They operate based on patterns in the data they have been trained on. If a novel vulnerability emerges after their training cut-off, they might not recognise it. Integration of language models into existing software development lifecycles requires careful consideration, especially concerning reliability, false positives/negatives, and the model's interpretability.

Looking ahead, there are multiple avenues for expanding upon this research:

\begin{enumerate}
    \item Broader SAST Tool Comparison: While our study focused on two specific SAST tools, future research could incorporate a broader range of tools to provide a more comprehensive comparison.
    \item Usability in Real-world Application: It would be beneficial to test GPT-4's detection capabilities in real-world scenarios, such as live software development environments, to assess its practical applicability and compare its output with expert opinion.
    \item Integration with Development Environments: Research could explore how GPT-4 or similar models can seamlessly and securely integrate into popular development environments and platforms.
    \item Security-trained LLMs: While GPT-4 is a generalised model, there might be benefits in training custom language models specifically focused on security vulnerability detection.
    \item LLMs-trained using Fault-Tolerant Quantum Computers(FTQC):
    \item Jens Eisert et al. \cite{eisert2023towards} provided a resource estimation for large machine learning models trained over Fault-Tolerant Quantum Computers (FTQC), focusing on significant computational expenses, power, and time consumption challenges. They demonstrated that FTQCs could offer efficient resolutions for generic (stochastic) gradient descent algorithms, scaling as \(O(T^2 \times \text{{polylog}}(n))\), where \(n\) is the size of the models and \(T\) is the number of iterations in training. The effectiveness depended on the models being sufficiently dissipative and sparse with minimal learning rates. The authors also explored the practical application, benchmarking models ranging from seven million to 103 million parameters, and found potential for quantum enhancement in sparse training after model pruning. This paper opens a new avenue for researching the resources and impact of training security-focused LLMs using FTQCs.
\end{enumerate}
 
Additionally, we must recognise that these advancements have both benefits and risks. While these models can help defence, attackers could also use them to find new vulnerabilities, introducing an asymmetry in Offensive Cyber Operations (OCO) that necessitates vigilant monitoring and research, which we introduce in the next section.

\subsection{Security concerns of LLMs}
CISA has emphasised that AI should adhere to the principle of "Secure by Design" \cite{CISA}, suggesting a comprehensive threat model tailored for domain-specific LLMs, such as GPTs specialised in vulnerability scanning. BSI has outlined several threats pertinent to AI security \cite{_2023_ai}.
Furthermore, it is crucial to recognise that many MLOps solutions rely on open-source frameworks.  This fact introduces heightened security vulnerabilities, especially concerning supply chain attacks on open-source resources and undetectable hidden backdoors \cite{goldwasser_2022_planting}. Shafi Goldwasser et al. shared an AI-era wisdom like \emph{Reflections on Trusting Trust}  \cite{thompson_1984_reflections}, which showcases undetectable backdoors in AI.

We outline the various attacks linked to LLMs; however, creating a comprehensive threat map for LLM-enhanced SAST is an important area for future research:
\begin{enumerate}
 \item \textbf{Poisoning Attacks:} Attackers introduce malicious data into the training set to compromise the model's performance \cite{Truong_2020_poisioning}. 
 \item \textbf{Backdoor Attacks:} Attackers embed a hidden behaviour within a model, triggered by specific inputs during deployment \cite{chen_2017_backdoor}. 
 \item \textbf{Supply chain attack:} Malicious activities aimed at tampering with the AI software supply chain to compromise the model or system \cite{williams2022discovery}. 
 \item \textbf{Endpoint /API security breach:} Exploiting vulnerabilities in the AI system's access points or interfaces to gain unauthorised access or leak information \cite{korolov_2023_why}. 
 \item \textbf{Model Stealing Attacks:} For organisations that invested significant resources in developing a commercial or mission-critical AI model, model stealing is a threat \cite{a2023_i}. 
 \item \textbf{Membership Inference Attacks:} In membership inference attacks, the attacker tries to determine whether a data sample was part of a model's training data \cite{rezashokri_2017_membership}. 
 \item \textbf{Attribute Inference Attacks:} In attribute inference attacks, the attacker seeks to breach the confidentiality of the model's training data by determining the value of a sensitive attribute associated with a specific individual or identity in the training data \cite{zhang_the}. 
 \item \textbf{Model Inversion Attacks:} Model inversion attacks aim to recover features that characterise classes from the training data \cite{chawla_datafree}. 
 \item \textbf{Denial of Service:} Attackers overload or manipulate the AI system, rendering it non-operational or degrading its performance \cite{_2023_ai}. 
 \item \textbf{Prompt injection:} Manipulating the input prompts to mislead or control the output of models like GPT-4  \cite{perez2022ignore}. 
 \item \textbf{Jailbreaks:} Bypassing restrictions or controls put on language models to access broader or hidden functionalities \cite{shen_2023_do}. 
 \item \textbf{Privacy breach:} Exploiting the model to reveal sensitive or private information it might have been exposed to during training \cite{pan2020privacy}. 
 \item \textbf{GAN-based Attack:} Using Generative Adversarial Networks to confuse or mislead the target AI model into making incorrect predictions or classifications \cite{_2023_ai}. 
\end{enumerate}
In this paper, we do not elaborate on the attack surface of LLMs. Instead, our focus is to underscore the significance of adhering to principles such as security by design/default and privacy by design/default. We advocate for integrating \textbf{MLSecOps} and emphasise the application of defence-in-depth strategies, notably the Zero Trust Architecture. It is also imperative to consider specific requirements like the Software Bill of Materials (SBOM) and other best practices when leveraging LLM-enhanced vulnerability scanning. These considerations are not just recommended; they are indispensable. Also, GPT4 and code-LLMs may generate insecure codes that warn about over-reliance on LLMs \cite{asare_2023_copilot}.

As revealed in the Vulkan files \cite{timberg_2023_secret}, reconnaissance systems are an undeniable component of cyber warfare. Therefore, it's imperative to enhance the resilience and robustness of LLM-enhanced SASTs through Federated Learning (FL-LLM), as these systems will become targets if they are not already. However, introducing FL-LLMs might also present new security challenges \cite{chen_2023_federated}. While using FL-LLM for training on a European scale/transatlantic scale might be feasible, training using a \emph{reliable dataset} is crucial as the dataset's quality will directly reflect the final performance of the model. Model hyperparameter tuning, retraining, and pruning will require substantial resources. Therefore, developing a high-quality European or even transatlantic dataset is unavoidable. Without such datasets, LLMs risk becoming costly failures due to the resource-intensive nature of training, model serving, and inference, leading to the potential for undertrained or poisoned models in cyber defence \cite{xue_2023,hoffmann_2022_training}. Present datasets \cite{chen_2023_hong} often lack comprehensive coverage of all known CWEs, proper labelling, and multi-language data.

\section{Conclusions}
This paper provides an empirical and statistically grounded comparison of a general-purpose LLM (GPT-4 Advanced Data Analysis) and two SAST tools for detecting security vulnerabilities in code snippets representing 32 common security pitfalls. Our primary scientific contribution is a transparent evaluation procedure that defines (i) a curated set of representative vulnerability scenarios, (ii) a consistent binary scoring rule for detection outcomes, (iii) an aggregated SAST baseline, and (iv) paired statistical inference using McNemar's test to compare detectors on the same samples. Within this controlled setting, GPT-4 achieved 30 correct detections out of 32 scenarios (93.75\%), and the paired comparison indicates a statistically significant performance difference in favour of GPT-4.

From an applicability perspective, the results support using LLM-assisted analysis as a complementary capability to SAST in secure software development lifecycles. In particular, LLM-based reasoning can help surface higher-level vulnerability mechanisms and contextual explanations that can accelerate triage and remediation, while SAST remains valuable for systematic rule-based analysis and integration into CI/CD workflows. The study also highlights that adopting LLM-enhanced scanning must be coupled with secure deployment practices, including careful handling of sensitive code, access controls, logging hygiene, and governance over model usage in development environments.

Limitations of this study include the bounded scenario set (32 samples), the use of only two SAST tools, reliance on a specific GPT-4 interface configuration at the time of testing, and the use of static code snippets rather than full projects and runtime execution. Scoring decisions, while documented for replication, may also be affected by detector reporting style and by borderline cases where a finding is related but not identical to the intended label. Future work will expand the scenario set and language coverage, evaluate additional SAST tools and LLMs, add expert adjudication to reduce scoring ambiguity, and incorporate project-level and runtime evaluations to better characterise false positives, false negatives, and operational impact in real DevSecOps pipelines.

\bibliography{main}

\section{Appendix}

\noindent
\textbf{Algorithm A1: McNemar test procedure for comparing GPT-4 and SAST}

\begin{enumerate}
    \item \textbf{Input:} A dataset \(D\) with one row per security scenario and binary columns:
    \begin{itemize}
        \item \texttt{GPT-Correct} \(\in \{0,1\}\)
        \item \texttt{SAST-Correct} \(\in \{0,1\}\)
    \end{itemize}
    \item \textbf{Derive paired disagreement counts:}
    \begin{itemize}
        \item \(b \leftarrow \left|\{i \in D : \texttt{SAST-Correct}_i = 1 \land \texttt{GPT-Correct}_i = 0\}\right|\) \hfill (SAST-only)
        \item \(c \leftarrow \left|\{i \in D : \texttt{SAST-Correct}_i = 0 \land \texttt{GPT-Correct}_i = 1\}\right|\) \hfill (GPT-only)
    \end{itemize}
    
    \item \textbf{Construct the 2x2 paired table:}
    \[
    \begin{bmatrix}
    a & b \\
    c & d
    \end{bmatrix}
    \quad \text{where } a \text{ and } d \text{ are not required for McNemar's statistic.}
    \]
    \item \textbf{Compute McNemar's test statistic:}
    \[
    \chi^2 \leftarrow \frac{(|b-c|-1)^2}{b+c}
    \]
    where the \(-1\) is the continuity correction and is used when \(b+c > 0\).
    \item \textbf{Compute p-value:} obtain \(p\) from a chi-square distribution with 1 degree of freedom using \(\chi^2\).
    \item \textbf{Decision rule:} if \(p < 0.05\), reject \(H_0\); otherwise, fail to reject \(H_0\).
    \item \textbf{Output:} \(\chi^2\), \(p\), and the hypothesis decision.
\end{enumerate}

\end{document}